\documentclass[aps,prx,twocolumn,nofootinbib,longbibliography]{revtex4-1}

\usepackage{subfig,braket,slashed,url,multirow,amsmath,amsfonts,xcolor}
\usepackage[T1]{fontenc} 
\allowdisplaybreaks[1]

\definecolor{myblue}{rgb}{0.33,0.30,0.85}
\definecolor{myred}{rgb}{1.02,0,0.05}

\usepackage[
colorlinks=true, 
pdfstartview=FitV, 
linkcolor=myblue, 
citecolor=myred,
urlcolor=myblue, 
bookmarks=true,
bookmarksnumbered=true,
pdftitle={},
pdfauthor={}
]{hyperref}

\usepackage[normalem]{ulem}

\newcommand{\ra}{\rightarrow} 
 
\newcommand{\lra}{\leftrightarrow}

\newcommand{\hyp}{\mathchar`-}

\newcommand{\U}{\mathrm{U}}

\newcommand{\bx}{{\mathbf x}}

\newcommand{\Z}{{\mathbb Z}}

\newcommand{\cO}{{\mathcal O}}
\newcommand{\cP}{{\mathcal P}}

\newcommand{\cS}{{\mathcal S}}

\newcommand{\non}{\nonumber}

\newcommand{\diff}{\mathrm{d}} 
\newcommand{\rmi}{\mathrm{i}} 
\newcommand{\rme}{\mathrm{e}} 

\newcommand{\edge}{\mathrm{edge}}

\begin{document}

\title{Anomaly-induced edge currents in hydrodynamics with parity anomaly}

\author{Takuya Furusawa}
\email{furusawa[at]stat.phys.titech.ac.jp}
\affiliation{Department of Physics, Tokyo Institute of Technology, Ookayama, Meguro, Tokyo 152-8551, Japan}

\author{Masaru Hongo}
\email{hongo[at]uic.edu}
\affiliation{Department of Physics, University of Illinois, Chicago, Illinois 60607, USA}
\affiliation{RIKEN iTHEMS, RIKEN, Wako 351-0198, Japan}

\date{\today}

\begin{abstract}
In this paper, we discuss relativistic hydrodynamics for a massless Dirac fermion in $(2+1)$ dimensions, which has the parity anomaly -- a global 't Hooft anomaly between $\U(1)$ and parity symmetries.
We investigate how hydrodynamics implements the party anomaly, particularly focusing on the transport phenomena at the boundary.
Based on the parity anomaly matching and the second law of local thermodynamics, 
we find $\U(1)$ and entropy currents localized at the boundary as well as the bulk anomalous current with vanishing divergence.
These edge currents are similar to the $(1+1)$-dimensional chiral transports, but the coefficients are given by half of theirs.
We also generalize our discussion to more general anomalies among multiple $\U(1)$ symmetries and single $\Z_2$ symmetry.
\end{abstract}

\maketitle

\section{Introduction}~\label{sec: intro}
An 't~Hooft anomaly represents an obstacle to introducing a background gauge field for a global symmetry of a quantum field theory.
The anomaly matching argument states that the 't Hooft anomaly is invariant under the renormalization group (RG) transformation and provides a nontrivial constraint on symmetry structures in the ultraviolet and infrared scales~\cite{tHooft:1980, Frishman:1980dq, Coleman:1982yg}.
In particular, the anomaly-matching constraint excludes possibilities of a unique gapped ground state at zero temperature. 
Therefore, the system with the 't Hooft anomaly must show
a nontrivial realization of the corresponding symmetry.
The well-known example of the 't~Hooft anomalies is the chiral anomaly, which partially explains why the chiral symmetry breaking occurs in quantum chromodynamics through the anomaly matching argument~\cite{tHooft:1980, Frishman:1980dq, Coleman:1982yg}.

Recent significant developments on the 't~Hooft anomaly are based on a detailed understanding of global symmetries and symmetry-protected topological (SPT) phases.
The former allows us to discuss subtle global anomalies involving discrete~\cite{Kapustin:2014lwa} and higher-form symmetries~\cite{Gaiotto:2014kfa}.
These new anomalies impose nontrivial constraints on low-energy behaviors of various systems in high-energy~\cite{Gaiotto:2017yup, Tanizaki:2017bam,Shimizu:2017asf, Yamazaki:2017dra, Tanizaki:2017mtm, Guo:2017xex, Cordova:2017kue,Tanizaki:2018wtg, Wan:2018zql, Yonekura:2019vyz, Cordova:2019jnf, Cordova:2019uob, Cordova:2019bsd, Hason:2020yqf, Cordova:2019wpi, Cordova:2019jqi, Furusawa:2020qdz, Honda:2020txe} and condensed matter physics~\cite{Wen:2013oza, Kapustin:2014tfa,Cho:2014jfa,Kapustin:2014gma, Kapustin:2014lwa, Wang:2014pma, Furuya:2015coa, Tachikawa:2016cha, Thorngren:2018wwt, Jian:2017skd, Cho:2017fgz, Komargodski:2017smk, Metlitski:2017fmd,  Sulejmanpasic:2018upi, Tanizaki:2018xto, Komargodski:2017dmc, Yao:2018kel,Wen:2018zux, Furusawa:2020kro}. 
On the other hand, a theory with an 't~Hooft anomaly emerges on a boundary of the corresponding SPT phase, which is described by a topological field theory in higher spacetime dimensions.
Their relation is known as the bulk-edge correspondence: the topological action of the bulk SPT phase inevitably links to the 't~Hooft anomaly of the boundary theory and vice versa~\cite{Wen:2013oza, Cho:2014jfa, Kapustin:2014lwa, Wang:2014pma}.

In addition to the above formal development, it is demonstrated that chiral anomalies survive in equilibrium and nonequilibrium settings, yielding peculiar transport phenomena present in both hydrodynamic and kinetic theory regimes. 
Well-known examples of the transport phenomena are the chiral magnetic effect (CME)~\cite{Vilenkin:1980fu,
Nielsen:1983rb, Alekseev:1998ds, Kharzeev:2007jp, Fukushima:2008xe} and chiral vortical effect (CVE)~\cite{Vilenkin:1979ui,Vilenkin:1980zv,Erdmenger:2008rm,Banerjee:2008th,Son:2009tf,Landsteiner:2011cp,Landsteiner:2011iq,Chen:2014cla} induced by the chiral anomaly. 
The CME/CVE represents the parity-breaking nondissipative currents along a magnetic field/vorticity.
These chiral transport phenomena have attracted much attention in high-energy, condensed matter, and nonequilibrium physics because they take place in diverse systems: 
a quark-gluon plasma created in heavy-ion collisions~\cite{Kharzeev:2007jp,Kharzeev:2010gd,Burnier:2011bf}, 
Weyl semimetals~\cite{Basar:2013iaa,Landsteiner:2013sja}, 
Floquet systems~\cite{Higashikawa:2018qsh,Sun:2018ads}, and 
non-Hermitian systems~\cite{Chernodub:2019ggz,Bessho:2020hrs,Kawabata:2020olo}.

The ubiquitous nature of the chiral transports comes from the RG-invariant property of the 't Hooft anomaly.
Hence, the global anomalies are also expected to cause intriguing transport phenomena analogous to the CME and CVE.
Nevertheless, in contrast to the chiral anomaly containing only the continuous chiral symmetry, a global anomaly often involves a discrete symmetry, so that the associated transports are unclear and have not been studied in detail until recently. 

In this paper, we investigate relativistic hydrodynamics consistent with the parity anomaly of the $(2+1)$-dimensional massless Dirac fermion~\cite{Niemi:1983rq, Redlich:1983kn, Redlich:1983dv, AlvarezGaume:1984zst} to take a closer look at transport phenomena induced by global anomalies.
We mainly focus on the system with a boundary, while the anomalous hydrodynamics with no boundary is derived from the kinetic theory~\cite{Chen:2013dca} and studied recently from the SPT viewpoint~\cite{Poovuttikul:2021mxx}.
We emphasize that consideration of the boundary leads to peculiar edge transports and is practically inevitable in laboratory experiments.

As we will show, the second law of local thermodynamics combined with the parity anomaly in the $(2+1)$-dimensional bulk requires nontrivial entropy and $\U(1)$ currents localized on the spatial boundary, whose forms are analogous to those of the chiral anomaly in $(1+1)$ dimensions~\cite{Loganayagam:2011mu, Dubovsky:2011sk, Jain:2012rh, Valle:2012em}.
Besides, we extend the discussion to discrete anomalies among multiple $\U(1)$ symmetries and single $\Z_2$ symmetry.
The extension may be helpful to analyze systems with multiple massless Dirac fermions, such as the graphene~\cite{Semenoff:1984dq}, and the $\mathbb{CP}^1$ model, an effective field theory of quantum antiferromagnets~\cite{Read:1989zz, Read:1990zza}.
We expect our analysis paves the way to formulate hydrodynamics with more generic global anomalies and shed light on an important relationship between the bulk and edge from the thermodynamic viewpoint.

This paper is organized as follows.
In Sec.~\ref{sec: preliminaries}, we review the parity anomaly and study the associated anomalous hydrodynamics in the absence of the boundary from the viewpoint of the anomaly matching.
In Sec.~\ref{sec: boundary}, we consider the case with a boundary and show that the consistency with the second law of local thermodynamics requests the presence of the nontrivial $\U(1)$ and entropy currents along the boundary.
We then generalize the anomalous hydrodynamics to the more general class of global anomalies in Sec.~\ref{sec: generalization}.
We summarize our results in the last section (Sec.~\ref{sec: summary}).

Throughout this paper, we consider systems
in the $(2+1)$-dimensional Minkowski spacetime $\mathbb{R}\times \Sigma$.
Here, $\Sigma$ is the spatial manifold, whose boundary is $\partial\Sigma$.
The spacetime indices are $\mu,\nu, \cdots = 0,1,2$, and 
the spacetime coordinate are referred to as $x^\mu = (t, \bx)$ ($t\in\mathbb{R}$, $\bx\in\Sigma$).
We employ the mostly plus convention for the metric 
$\eta_{\mu\nu} = \mathrm{diag}(-1,1,1)$ and the Levi-Civita symbol
$\epsilon^{\mu\nu\rho}$ with $\epsilon^{012} = +1$.

\section{Preliminaries}~\label{sec: preliminaries}
In this section, we briefly review the parity anomaly and the associated anomalous hydrodynamics in the absence of the boundary (i.e., $\partial\Sigma = 0$) from the anomaly-matching viewpoint.
It turns out that the discussion here is almost parallel to the nonanomalous hydrodynamics except for an anomalous contribution to the $\U(1)$ current as studied in Refs.~\cite{Chen:2013dca, Poovuttikul:2021mxx}.%
\footnote{See also Ref.~\cite{Jensen:2011xb} for a general parity-violating hydrodynamics in $(2+1)$ dimensions.}

\subsection{Parity anomaly}~\label{sec: parity anomaly}
Let us consider the Dirac fermion in $(2+1)$ dimensions.
The Dirac mass in $(2+1)$ dimensions explicitly breaks the parity symmetry, and the massless Dirac fermion is parity symmetric classically.
However, the symmetry is anomalous, i.e., it is spoiled by quantum effects
when we introduce the background $\U(1)$ gauge field.
This quantum violation of the parity symmetry is known as the parity anomaly~\cite{Niemi:1983rq, Redlich:1983kn, Redlich:1983dv, AlvarezGaume:1984zst}.

The violation is manifest if one takes the Pauli-Villars regularization:
i.e., the cutoff parameter incorporated as the Dirac mass of the Pauli-Villars regulator breaks the parity symmetry.
In the presence of the background $\U(1)$ gauge field $A_\mu(x)$,
the parity-violating contribution to the effective action is given by~\cite{Niemi:1983rq,Redlich:1983kn,Redlich:1983dv}
\begin{equation}~\label{eq: Pauli Villars}
    \begin{split}
        S_\mathrm{anom}[A] &= -\eta[A]/2 \\
        &=-\int
        d^3x \left(
            \frac{1}{8 \pi}\epsilon^{\mu\nu\rho}A_\mu (x)\partial_\nu A_\rho(x)
        \right)+\cdots,
    \end{split}
\end{equation}
where $\eta[A]$ is the Atiyah-Patodi-Singer $\eta$ invariant~\cite{Atiyah:1975jf}.
In the second line, we expanded $\eta[A]$ perturbatively.

We shall rephrase the parity anomaly in a more formal language for later convenience.
Due to the background gauge field $A_\mu(x)$, the partition function of the massless Dirac fermion is not invariant under the parity transformation and varies as 
\begin{equation}~\label{eq: anomalous partition function}
    Z[\cP A, \cP \chi] = Z[A,\chi] \rme^{\rmi \eta[A]},
\end{equation}
with background sources $\chi_a(x)$ for arbitrary local operators $\cO^a(x)$.
Here, we defined $[\cP A]_\mu(x) = P^\nu_{\mu} A_\nu(\cP x)$ and $[\cP x]^\mu =P^{\mu}_\nu x^\nu $ with $P^\mu_\nu = \mathrm{diag}(1,1,-1)$.
The transformation rules for $\chi_a(x)$ are appropriately determined depending on those for $\cO^a(x)$ so that the action is parity invariant.
The phase factor on the right-hand side results from the contribution of, e.g., the Pauli-Villars regulator and is the signal of the parity anomaly.

Differentiating Eq.~\eqref{eq: anomalous partition function} with respect to $A_\mu(x)$ yields 
\begin{equation}~\label{eq: average anomalous current}
    \braket{\cP J^\mu(x)}_{\cP A,\cP \chi} = \braket{J^\mu(x)}_{A,\chi}  + \frac{1}{2\pi}\epsilon^{\mu\nu\rho}\partial_\nu A_\rho(x),
\end{equation}
where $\cP J^\mu(x) = P^\mu_{\nu} J^\nu(\cP x)$, and $\braket{\cdots}_{A,\chi}$ is the average in the presence of the background fields.
This equation indicates that $\braket{J^\mu(x)}_{A,\chi}$ 
and the parity transformation of $\braket{J^\mu(x)}_{\cP A,\cP \chi} $ 
differ by the Hall current coming from $\eta[A]$.
Thanks to the background sources $\chi_a(x)$ in Eq.~\eqref{eq: anomalous partition function}, we can show Eq.~\eqref{eq: average anomalous current} in the presence of arbitrary operators $\cO^a(x)$.
Therefore, it is a property of the $\U(1)$ current operator rather than the ground state.
Consequently, Eq.~\eqref{eq: average anomalous current} must survive even in a nonequilibrium regime, and it gives a nontrivial constraint on the anomalous hydrodynamics of the massless Dirac fermion.

\subsection{Anomalous hydrodynamics without boundary}
We then study how hydrodynamics incorporates the parity anomaly in the absence of the boundary from the perspective of anomaly matching.
First of all, the time evolution of relativistic hydrodynamics is governed by the conservation laws of the energy-momentum tensor $T^{\mu\nu}(x)$ and the $\U(1)$ current $J^\mu(x)$:
\begin{subequations}
~\label{eq: conservation laws w/o boundary}
 \begin{align}
  \partial_\mu T^{\mu\nu}(x) &= F^{\nu\rho}(x)J_\rho(x),
  ~\label{eq: energy momentum conservation w/o boundary} \\
  \partial_\mu J^\mu(x) &=0.
  ~\label{eq: particle conservation w/o boundary}
 \end{align}    
\end{subequations}
Here, $F_{\mu\nu}(x) = \partial_\mu A_\nu(x) - \partial_\nu A_\mu(x)$ is the field strength of $A_\mu(x)$.

The dynamical variables of hydrodynamics are given by the energy density $e (x)$, charge density $n(x)$, and normalized fluid velocity $u^\mu(x)$ with $u^\mu(x)u_\mu(x) = -1$.
Throughout this paper, we use the Landau-Lifshitz frame for the fluid velocity~\cite{Landau:1987Fluid}, 
and thus the fluid velocity satisfies
\begin{equation}
    T^\mu_{~\nu} (x) u^\nu (x) = - e (x) u^\mu (x).~\label{eq: Landau-Lifshitz condition}
\end{equation}

We also use local thermodynamic variables $\beta (x) = 1/T(x)$ and $\nu (x) = \beta(x) \mu (x)$, where $T(x)$ and $\mu(x)$ are the local temperature and chemical potential, respectively.
These variables are conjugate to $e(x)$ and $n(x)$ and defined as the derivatives of the entropy density $s(x) \equiv s (e(x),n(x))$:
\begin{subequations}
    \begin{align}
        \beta (x) &\equiv \frac{\partial s (x) }{\partial e (x)} ,
        \\
        \nu (x) &\equiv - \frac{\partial s (x) }{\partial n (x)}. 
    \end{align}
    ~\label{eq:def-conjugate-variables}
\end{subequations}
Note that the above definitions~\eqref{eq:def-conjugate-variables} are equivalent to the first law of local thermodynamics: 
\begin{equation}
    T (x) \diff s (x) = \diff e (x) - \mu (x) \diff n(x).\label{eq:first-law0}
\end{equation}
As we will show, the second law of local thermodynamics leads to the Gibbs-Duhem relation:
\begin{equation}
    T (x) s (x) + \mu (x) n(x) = e(x) + p(x),
    \label{eq:Gibbs-Duhem}
\end{equation}
where $p (x)$ denotes the pressure of the fluid.
With the help of this, one can also express the first law~\eqref{eq:first-law0}
in the following conjugate form:
\begin{equation}
    \big[e(x) + p(x) \big] \diff \beta (x) 
    + \beta (x) \diff p (x) = n (x) \diff \nu (x).~\label{eq: first law}
\end{equation}

The conservation laws \eqref{eq: conservation laws w/o boundary} are closed once we find the constitutive relations.
For that purpose, we rely on the derivative expansion.
We employ the power-counting scheme that counts 
all the dynamical variables $\beta(x)$, $\nu(x)$ [or $e(x),n(x)$], $u^\mu(x)$, and the background gauge field $A_\mu(x)$
as zeroth-order quantities with respect to the derivative, i.e.,
\begin{equation}
 \{\beta(x), \nu(x), u^\mu (x), A_\mu (x)\} = \cO(\partial^0).
\end{equation}
We then expand the energy-momentum tensor and $\U(1)$ current with respect to the derivative as 
\begin{subequations}
    \begin{align}
        T^{\mu\nu} (x) =& T^{\mu\nu}_{(0)} (x) + T^{\mu\nu}_{(1)}(x)+\cO(\partial^2), ~\label{eq: energy momentum tensor}
        \\
        J^{\mu} (x) =& J^{\mu}_{(0)} (x)+J^{\mu}_{(1)} (x)+J^{\mu}_\mathrm{anom} (x)+\cO(\partial^2),
        ~\label{eq: particle current}
    \end{align}
\end{subequations}
where $T^{\mu\nu}_{(n)}(x)$ and $J^\mu_{(n)}(x)$ are $\cO(\partial^n)$ terms, and $J^\mu_\mathrm{anom}(x)$ is an anomalous current required from Eq.~\eqref{eq: average anomalous current}.

The leading order contributions, $T^{\mu\nu}_{(0)}(x)$ and $J^\mu_{(0)}(x)$ are given by those of the relativistic perfect fluid:
\begin{subequations}
    \begin{align}
        T^{\mu\nu}_{(0)} (x) = & [e(x)+p(x)] u^\mu(x) u^\nu(x) + p(x) \eta^{\mu\nu},
        \\
        J^{\mu}_{(0)} (x) = & n(x) u^\mu(x).
    \end{align}
\end{subequations}
These forms are determined from the condition that $T^{\mu\nu}(x)$ and $J^\mu(x)$ are related to the thermodynamic quantities in the local rest frame (LRF), where $u^\mu(x) \big|_{\mathrm{LRF}} =(1,0,0)$.
More precisely, we demand that they satisfy
$T^{\mu\nu}(x) \big|_{\mathrm{LRF}} = \mathrm{diag}(e(x),p(x),p(x))$ and $J^\mu(x)-J^\mu_\mathrm{anom}(x) \big|_{\mathrm{LRF}} = (n(x),0,0)$.
Note that we adopt the definition of $n (x)$ as 
$ n (x) \equiv - u_\mu(x) (J^\mu(x) - J^\mu_\mathrm{anom}(x))$ 
to simplify the following discussion.%
\footnote{
Another familiar choice is to define
the particle number density as $u_\mu(x) J^\mu(x) = -n(x)$~\cite{Chen:2013dca,Poovuttikul:2021mxx}.
Since $u_\mu(x)J^\mu_\mathrm{anom}(x) = C B(x)$, we can readily rephrase our hydrodynamic equations by shifting the thermodynamic variables as
$n(x) \ra n(x)+ CB(x)$,
$p(x) \ra p(x)  + (\partial p/\partial n)_e CB(x)$,
$e(x) \ra e(x)$,
$\mu(x) \ra \mu(x)  + (\partial \mu/\partial n)_e CB(x)$,
$T(x) \ra T(x)  + (\partial T/\partial n)_e CB(x)$,
and
$s(x) \ra s(x)-\nu(x) CB(x)$.
Here, we differentiated the thermodynamic variables fixing $e(x)$ to keep the condition $u_\mu(x)u_\nu(x) T^{\mu\nu}(x) = e(x)$ and used the thermodynamic equations~\eqref{eq:def-conjugate-variables} to show $(\partial s/ \partial n)_e = - \nu(x)$.}
We also note that the first-order contributions satisfy $u_\mu(x) T^{\mu\nu}_{(1)}(x) = 0$ and $u_\mu(x) J^\mu_{(1)}(x) = 0$ because of our definition of $e(x)$ and $n(x)$ in the Landau-Lifshitz frame.
The forms of $T^{\mu\nu}_{(1)}(x)$ and $J^{\mu}_{(1)}(x)$ are determined from the second law of thermodynamics, as explained below.

To reproduce the anomalous property of the $\U(1)$ current~\eqref{eq: average anomalous current}, Eq.~\eqref{eq: particle current} needs to be equipped with the anomalous current $J^{\mu}_\mathrm{anom} (x)$ given by
\begin{equation}
    \begin{split}~\label{eq: anomalous current}
        J^{\mu}_\mathrm{anom} (x) 
        =& C \epsilon^{\mu\nu\rho} \partial_\nu A_\rho(x)
        \\
        =& - C u^\mu(x)B(x) + C\epsilon^{\mu\nu\rho} u_\nu(x)E_\rho(x),
    \end{split}
\end{equation}
with the constant $C = - 1/(4 \pi)$.
In the second line, we defined the magnetic field $B(x) = \epsilon^{\mu\nu\rho}u_\mu(x)\partial_\nu A_\rho(x)$ and electric field $E_\mu(x) = F_{\mu\nu}(x) u^\nu(x)$.
This Hall current reproduces the equation~\eqref{eq: average anomalous current} as follows:
\begin{equation}~\label{eq: parity transformation}
    \begin{split}
        \cP J^{\mu}_\mathrm{anom}(x) \Big|_{\cP A}
        =& C [\epsilon^{\mu'\nu'\rho'} P^\mu_{\mu'} P^\nu_{\nu'} P^\rho_{\rho'}] \partial_{\nu} A_{\rho}(x)
        \\
        =& J^{\mu}_\mathrm{anom}(x) \Big|_{A}+ \frac{1}{2\pi} \epsilon^{\mu\nu\rho} \partial_\nu A_\rho(x).
    \end{split}
\end{equation}
Here, $J^{\mu}_\mathrm{anom} \Big|_{A}$ represents the anomalous current~\eqref{eq: anomalous current} computed with the background field $A_\mu(x)$.
Hence, we have 
$J^{\mu}_\mathrm{anom}(x) \Big|_{\cP A} = -\frac{1}{4\pi} \epsilon^{\mu\nu\rho'} P^{\rho}_{\rho'}  \partial_\nu A_{\rho}(\cP x)$.
The parity transformation of this equation proves the first line in Eq.~\eqref{eq: parity transformation}.
Note that we have also used the identity $\epsilon^{\mu'\nu'\rho'} P^\mu_{\mu'} P^\nu_{\nu'} P^\rho_{\rho'} = -\epsilon^{\mu\nu\rho}$ in the second line.

We emphasize that the anomalous current does not affect the conservation law of $J^\mu(x)$ because $J^{\mu}_\mathrm{anom} (x)$ identically satisfies $\partial_\mu J^{\mu}_\mathrm{anom} (x) = 0$.
Hence, in contrast to the case with the chiral anomaly~\cite{Son:2009tf}, the hydrodynamics with the parity anomaly is described by the same conservation laws as the standard relativistic hydrodynamics~\cite{Landau:1987Fluid}.

We can also specify the constitutive relations for $T^{\mu\nu}_{(1)}(x)$ and $J^\mu_{(1)}(x)$ by requiring the second law of local thermodynamics:
\begin{equation}~\label{eq: second law w/o boundary}
    \partial_\mu s^\mu(x) \ge 0,
\end{equation}
with the entropy current $s^\mu(x)$.
Due to $\partial_\mu J^{\mu}_\mathrm{anom} (x) = 0$,
one can introduce the entropy current in the standard way:
\begin{equation}
    s^\mu(x) = s(x) u^\mu(x) - \nu(x)J^\mu_{(1)}(x),
\end{equation}
whose divergence is readily computed from Eq.~\eqref{eq: energy momentum conservation w/o boundary} and Eq.~\eqref{eq: particle conservation w/o boundary} and given by
\begin{align}
    \partial_\mu s^\mu(x)
    =& - \big[ s(x) - \beta(x) \big( e (x) + p (x)- \mu (x) n(x) \big) \big] \partial_\mu u^\mu(x)
    \non \\
    &- T^{\mu\nu}_{(1)}(x)\frac{\partial_\mu u_\nu(x) }{T(x)}
    \non \\
    &- J^\mu_{(1)}(x) \left[\partial_\mu \nu (x)- \frac{E_\mu(x)}{T(x)}\right] .
 \label{eq: entropy production w/o boundary}
\end{align}
Thus, from the requirement of the second law~\eqref{eq: second law w/o boundary}, we derive the Gibbs-Duhem relation~\eqref{eq:Gibbs-Duhem}%
\footnote{
One may find the possible derivative correction to the Gibbs-Duhem relation as
\begin{equation}
 T(x)s(x) - e (x) + p (x)- \mu (x) n(x) = -\alpha(x) \partial_\mu u^\mu,
\end{equation}
with a positive coefficient function $\alpha(x)$.
However, we can eliminate this correction by redefining the bulk viscosity $\zeta$.
}
and the following constitutive relations: 
\begin{subequations}
    \begin{align}
        T^{\mu\nu}_{(1)}(x) =& - \eta (x)\Delta^{\mu\alpha}(x)\Delta^{\nu\beta}(x)\left[\partial_\alpha u_\beta(x)+\partial_\beta u_\alpha(x)\right] \non
        \\
        &-[\zeta(x) - \eta(x)] \Delta^{\mu\nu}(x)\partial_\alpha u^\alpha(x),~\label{eq: T1}
        \\
        J^\mu_{(1)}(x) =& - T(x) \sigma(x) \Delta^{\mu\nu}(x) \left[ \partial_\nu \nu(x)-  \frac{E_\nu(x)}{T(x)}\right] ,~\label{eq: J1}
 \end{align}
\end{subequations}
with $\Delta^{\mu\nu}(x) = \eta^{\mu\nu} + u^\mu(x)u^\nu(x)$ 
and three positive functions of the thermodynamic variables, $\eta(x)$, $\zeta(x)$, and $\sigma(x)$.
These three coefficients are known as the shear viscosity, bulk viscosity, and charge conductivity, respectively~\cite{Landau:1987Fluid}.

\section{Boundary Entropy Production}~\label{sec: boundary}

Following the analysis given in the previous section,
we shall study how the presence of the
boundary affects the transport phenomena. 
The essential difference is that the boundary modifies the conservation laws~\eqref{eq: conservation laws w/o boundary} as
\begin{subequations}~\label{eq: conservation laws w/ boundary}
    \begin{align}
        \partial_\mu T^{\mu\nu}(x) 
        + \partial_\mu\Theta(\bx \in \Sigma) T^{\mu\nu}(x)
        &= F^{\nu\rho}(x)J_\rho(x),
        ~\label{eq: energy momentum conservation w/ boundary}
        \\
        \partial_\mu J^\mu(x) + \partial_\mu\Theta(\bx \in \Sigma)J^{\mu}(x)
        &=0.
        ~\label{eq: particle conservation w/ boundary}
    \end{align}    
\end{subequations}
Here, we defined the Heaviside step function $\Theta(\bx \in \Sigma)$, which takes one for $\bx\in\Sigma$ and otherwise vanishes. 
We can write $\partial_\mu \Theta(\bx \in \Theta)$ as $- N_\mu(x)\delta(\bx \in \partial\Sigma)$ using the normal vector $N_\mu(x)$ on $\partial\Sigma$ with $N_\mu(x)N^\mu(x) = 1$ and the delta function $\delta(\bx \in \partial\Sigma)$, which is nonzero only at the boundary $\bx \in \partial\Sigma$.
The boundary terms in Eqs.~\eqref{eq: conservation laws w/ boundary} subtract the currents flowing out of the bulk $\Sigma$.%
\footnote{
One can understand the boundary contributions as follows.
Suppose a generic system with a dynamical field $\phi(x)$ 
has a $\U(1)$ symmetry as $\delta_{\theta} S [\phi] = 0$, where $\delta_\theta$ denotes the infinitesimal $\U(1)$ transformation.
Promoting $\theta$ to the local function $\theta(x)$, 
we find the induced variation of the action $S[\phi]$ under the $\U(1)$ gauge transformation takes the form:
\begin{equation}
    \begin{split}
        \delta_{\theta} S[\phi] = \int_{\mathbb{R}\times\Sigma} d^3x \partial_\mu \theta(x&) J^\mu(x)\\
        = - \int_{\mathbb{R}^3} d^3x \theta(x) 
        \Big[&
        \Theta(\bx \in \Sigma)\partial_\mu J^\mu(x)
        \\
        &+\partial_\mu \Theta(\bx \in\Sigma) J^\mu(x)
        \Big].
    \end{split}
\end{equation}
Requiring $\phi(x)$ satisfies its equation of motion, we obtain the conservation law in the presence of the boundary as
\begin{equation}
    \partial_\mu J^\mu(x) +\partial_\mu \Theta(\bx \in\Sigma) J^\mu(x) = 0.
\end{equation}
}
Similarly, we need to change the second law and require
\begin{equation}~\label{eq: second law w/ boundary}
    \partial_\mu s^\mu(x) 
    +\partial _\mu\Theta(\bx \in \Sigma)s^\mu(x) 
    \ge 0.
\end{equation}

We then reexamine the second law~\eqref{eq: second law w/ boundary} using the modified conservation laws~\eqref{eq: conservation laws w/ boundary}.
Computing the left-hand side of Eq.~\eqref{eq: second law w/ boundary}, we find
\begin{equation}~\label{eq: entropy production w/ boundary}
    \begin{split}
        \partial_\mu& s^\mu(x)+\partial_\mu \Theta(\bx \in \Sigma) s^\mu(x)
        \\
        =&
        - T^{\mu\nu}_{(1)}(x)\frac{\partial_\mu u_\nu(x) }{T(x)}
        \\
        & -J^\mu_{(1)}(x) \left[ \partial_\mu \nu(x) - \frac{E_\mu(x)}{T(x)}\right] 
        \\ 
        &+ \partial_\mu\Theta(\bx \in \Sigma) C \nu(x) \epsilon^{\mu\nu\rho}u_\nu(x)E_\rho(x),
    \end{split}
\end{equation}
where we used the conservation laws~\eqref{eq: conservation laws w/ boundary}, 
the definition of conjugate variables~\eqref{eq:def-conjugate-variables}, the Gibbs-Duhem relation \eqref{eq:Gibbs-Duhem}, and the boundary condition $u^\mu(x)\partial_\mu\Theta(\bx\in\Sigma) = 0$.
This boundary condition means the fluid velocity is always perpendicular to the normal vector, and the energy current cannot go through the boundary.
The first two terms are always positive if we take Eq.~\eqref{eq: T1} and Eq.~\eqref{eq: J1}.
However, the last term can be negative and violate the second law at the boundary~\eqref{eq: second law w/ boundary}.
Note that the violation is of order $\cO(\partial^1)$ on $\partial \Sigma$.%
\footnote{We regard $\partial_\mu \Theta(\bx\in\Sigma) = -N_\mu(x)\delta(\bx\in\partial\Sigma)$ as $\cO(\partial^0)$ on the boundary $\partial \Sigma$.}

Hence, we have to introduce new terms to the currents to fix the violation of the second law at the boundary.
These terms must be $\cO(\partial^0)$ quantities on $\partial\Sigma$ to cancel the $\cO(\partial^1)$ violation, be perpendicular to $u^\mu(x)$ such as $T^{\mu\nu}_{(1)}(x)$ and $J^{\mu}_{(1)}(x)$, be parity odd, and vanish in the absence of the boundary.
These four requirements allow us to add only the following boundary contributions to the $\U(1)$ and entropy currents:
\begin{subequations}\label{eq:edge-current}
    \begin{align}
        J^\mu_\edge (x) =&  \xi (x)\partial_\nu\Theta(\bx\in\Sigma)\epsilon^{\mu\nu\rho}u_\rho(x),
        \\
        s^\mu_\edge(x) =& -\nu(x)  J^\mu_\edge (x) \non 
        \\
        &+ D (x) \partial_\nu\Theta(\bx\in\Sigma)\epsilon^{\mu\nu\rho}u_\rho(x),
    \end{align}
\end{subequations}
with $\xi(x)$ and $D(x)$, local functions of the thermodynamic variables.
We will show the local second law determines these coefficient functions up to a certain constant in the following.

Let us repeat the computation of the second law~\eqref{eq: second law w/ boundary} with the new currents.
Thanks to additional contributions from the edge currents~\eqref{eq:edge-current}, the entropy production now takes the form:
\begin{equation}
    \begin{split}
        \partial_\mu s^\mu(x) +&\partial_\mu \Theta(\bx\in\Sigma)s^\mu(x)
        \\
        = - T^{\mu\nu}_{(1)}&(x)\frac{\partial_\mu u_\nu(x) }{T(x)}
        -J^\mu_{(1)}(x)\left[\partial_\mu \nu (x) - \frac{E_\mu(x)}{T(x)}\right] 
        \\
        +\partial_\mu&\Theta(\bx \in \Sigma)\epsilon^{\mu\nu\rho}u_\nu(x)
        \\
        \times&\left[C \nu(x)+ \frac{n(x)D(x)}{e(x)+p(x)}+ \frac{\xi(x)}{ T(x)}\right] E_\rho(x)
        \\
        +\partial_\mu&\Theta(\bx \in \Sigma)\epsilon^{\mu\nu\rho}u_\nu(x)
        \\
        \times&\left[\partial_\rho D (x)- \frac{D(x)\partial_\rho p(x)}{e(x)+p(x)}- \xi(x) \partial_\rho \nu (x)\right].
    \end{split}
\end{equation}
Here, we have used the following equation derived from Eq.~\eqref{eq: energy momentum conservation w/ boundary} and the identity $\epsilon^{\mu\nu\rho}\partial_\mu \partial_\nu \Theta(\bx\in\Sigma) = 0$:
\begin{equation}
    \begin{split}
        &\partial_\mu \left[
            \epsilon^{\mu\nu\rho}\partial_\nu \Theta(\bx\in\Sigma)u_\rho(x)
        \right] 
        \\
        &=\epsilon^{\mu\nu\rho}\partial_\mu \Theta(\bx\in\Sigma)u_\nu(x) \frac{n(x)E_\rho(x)-\partial_\rho p(x) + \cO(\partial^2)}{e(x)+p(x)}.
    \end{split}
\end{equation}

Therefore, we recover the second law if the coefficients of the boundary currents satisfy the following relations:
\begin{subequations}
    \begin{align}
        C \nu(x)=& -\frac{n(x)D(x)}{e(x)+p(x)}-\frac{\xi(x)}{T(x)},~\label{eq: condition 1}
        \\
        \partial_\rho D (x)=& \frac{D(x)}{e(x)+p(x)} \partial_\rho p(x) +\xi(x)\partial_\rho \nu(x).~\label{eq: condition 2}
    \end{align}
\end{subequations}
We can solve these relations in the following way.
Regarding $D(x)$ and $p(x)$ as functions of $\beta(x)$ and $\nu(x)$,
we can rewrite the second equation~\eqref{eq: condition 2} as 
\begin{equation}
    \begin{split}
        0 =& \left( \frac{\partial D(x)}{\partial \beta (x)}+ T (x) D (x) \right) \partial_\rho \beta (x)
        \\
        & + \left( \frac{\partial D(x)}{\partial \nu (x)}- \frac{T (x) n (x) D (x)}{ e(x)+ p (x)} - \xi (x) \right) \partial_\rho \nu (x),
    \end{split}
\end{equation}
where we used the first law~\eqref{eq: first law}.
This equation holds identically when the following equations are satisfied:
\begin{subequations} 
    \begin{align}
        0&= \frac{\partial D(x)}{\partial \beta (x)}+ T (x) D (x),~\label{eq:condition-2-1}
        \\
        0&= \frac{\partial D(x)}{\partial \nu (x)}- \frac{T (x) n(x) D (x)}{ e(x)+ p (x)} - \xi (x) . ~\label{eq:condition-2-2}
    \end{align}
\end{subequations}
One can immediately find the solution of Eq.~\eqref{eq:condition-2-1} taking the form:
\begin{equation}
    D (x) = \beta^{-1} (x) g \big( \nu(x) \big) .
\end{equation}
Here, $g \big( \nu(x) \big)$ is a certain function depending only on $\nu(x)$.
Substituting this expression to Eq.~\eqref{eq:condition-2-2}, we obtain 
\begin{equation}
    \begin{split}
        0&= \frac{\diff g \big( \nu(x) \big)}{\diff \nu (x)} - \left[\frac{n(x) D (x) }{e(x)+p(x)}+\frac{\xi (x)}{T(x)}\right] 
        \\
        &= \frac{\diff g \big( \nu(x) \big)}{\diff \nu (x)} + C \nu(x),
    \end{split}
\end{equation}
where we used Eq.~\eqref{eq: condition 1} to get the second line.
We then solve this equation for $g \big( \nu (x) \big)$ and obtain the following expressions for $D (x)$ and $\xi (x)$:
\begin{subequations}
    \begin{align}
        D(x) &= -\frac{C\mu(x)^2}{2T(x)} - cT(x),
        \\
        \xi(x) & = -C\mu(x) + \frac{n(x)\left[\frac{C\mu(x)^2}{2} + cT(x)^2\right]}{e(x)+p(x)},
    \end{align}
\end{subequations}
with an integration constant $c$.

In summary, the parity anomaly induces the following $\U(1)$ and entropy currents localized at the boundary:
\begin{subequations}~\label{eq:anomalous-edge-currents}
    \begin{align}
        J^\mu_{\edge} (x) =-&C\mu(x) \epsilon^{\mu\nu\rho}\partial_\nu\Theta(\bx\in\Sigma)u_\rho(x) \non
        \\
        +& \frac{n(x)}{e(x)+p(x)}\left[\frac{C\mu(x)^2}{2} + cT(x)^2\right] \non 
        \\
        &\times \epsilon^{\mu\nu\rho}\partial_\nu\Theta(\bx\in\Sigma)u_\rho(x),~\label{eq: boundary particle current}
        \\
        s^\mu_{\edge} (x) = -& \nu(x)J^\mu_{\edge} (x) \non
        \\
        -& \left[\frac{C\mu(x)^2}{2T(x)} + cT(x)\right] \epsilon^{\mu\nu\rho}\partial_\nu\Theta(\bx\in\Sigma)u_\rho(x).~\label{eq: boundary entropy current}
    \end{align}
\end{subequations}
Therefore, in the system with the parity anomaly, the finite chemical potential and temperature induce the nonzero edge currents automatically.

The anomaly-induced edge currents in Eqs.~\eqref{eq:anomalous-edge-currents} take similar forms to $(1+1)$-dimensional analogs of the CME and CVE associated with the Weyl fermion~\cite{Loganayagam:2011mu, Dubovsky:2011sk, Jain:2012rh, Valle:2012em}.
However, we emphasize that the coefficient $C = -1/(4\pi)$ is half of the $(1+1)$-dimensional ones.
This observation is plausible because the Weyl fermion appears on the boundary of the Chern insulator, whose bulk SPT action is given by the $\eta$ invariant $\eta[A]$ and twice of Eq.~\eqref{eq: Pauli Villars}. 
We also note that the constant $c$, which we cannot determine from our thermodynamic consideration, may be determined from gravitational contributions to the $\eta$ invariant, but clarifying this point is left for future work.

\section{Generalization to $\U(1)^m\hyp\Z_2$ anomaly}~\label{sec: generalization}
Lastly, let us generalize the above discussion to systems with a mixed anomaly among $m$-multiple $\U(1)$ symmetries and a $\Z_2$ symmetry.
This generalization includes anomalies of spinless electrons on a honeycomb lattice and the easy-plane $\mathbb{CP}^1$ model for quantum antiferromagnets.

Suppose that the $\Z_2$ symmetry transformation $\cS$ acts on the partition function as
\begin{equation}
    Z[\cS A_a] = Z[A_a]\rme^{-\rmi \int \diff^3 x C_{ab}\epsilon^{\mu\nu\rho}A^a_\mu(x)\partial_\nu A^b_\mu(x)},~\label{eq: general anomaly}
\end{equation}
where we introduced an $\cS$-odd $m\times m$ symmetric matrix $C_{ab}$ and a set of the background gauge fields for the $\U(1)$ symmetries $\{A^a_\mu(x)\}^m_{a=1}$.
This class of anomalies often appears in $(2+1)$-dimensional systems.
First, Eq.~\eqref{eq: general anomaly} reduces to the parity anomaly~\eqref{eq: anomalous partition function} if we take $m = 1$ and $C_{11}=-1/(4\pi)$ and regard the $\Z_2$ symmetry as the parity symmetry.
Moreover, the case with $m = 2$ and $C_{ab} = -(\sigma_z)_{ab}/(4\pi)$
corresponds to spinless electrons on a honeycomb lattice [$\sigma_{\alpha}$ ($\alpha = x,y,z$) represents the Pauli matrix], where the valley degrees of freedom suffer from the parity anomalies~\cite{Semenoff:1984dq}.
Besides, the easy-plane $\mathbb{CP}^1$ model, an effective field theory for quantum antiferromagnets, has the $\mathrm{O}(2)_\mathrm{S}\times\U(1)_\mathrm{M}$ anomaly with the coefficient $C_{ab} = -(\sigma_x)_{ab}/(4\pi)$. (See Ref.~\cite{Wang:2017txt} for example.)

Recalling the discussion in Sec.~\ref{sec: preliminaries}, one can show this anomaly induces the following bulk anomalous current in hydrodynamics:
\begin{equation}
    J^\mu_{\textrm{anom},a}(x) = C_{ab}\epsilon^{\mu\nu\rho}\partial_\nu A^b_\mu(x).
\end{equation}
We then repeat the entropy-current analysis in the presence of the boundary and derive the following $\U(1)$ and entropy currents on the boundary:
\begin{subequations}
    \begin{align}
        J^\mu_{\edge,a}(x)=&-C_{ab}\mu^b(x)\epsilon^{\mu\nu\rho}\partial_\nu\Theta(\bx\in\Sigma)u_\rho(x),\non
        \\
        + &\frac{n_a(x)}{e(x)+p(x)}\left[\frac{C_{bc}\mu^b(x)\mu^c(x)}{2} + cT(x)^2\right] \non
        \\
        &\times \epsilon^{\mu\nu\rho}\partial_\nu\Theta(\bx\in\Sigma)u_\rho(x),
        \\
        s^\mu_{\edge}(x) = -& \frac{\mu^a(x)}{T(x)}J^\mu_{a,\edge}(x) \non
        \\
        - & \left[\frac{C_{ab}{\mu^a}(x){\mu^b}(x)}{2T(x)} + cT(x)\right] \non
        \\
        &\times\epsilon^{\mu\nu\rho}\partial_\nu\Theta(\bx\in\Sigma)u_\rho(x),
    \end{align}
\end{subequations}
with an undetermined constant $c$. 
Here, $n_a(x)$ and $\mu^a(x)$ ($a=1,\cdots,m$)
are the particle densities and chemical potentials associated with the $\U(1)$ symmetries, respectively.

\section{Summary and discussions}~\label{sec: summary}
In this paper, we studied nondissipative transports of relativistic hydrodynamics with the parity anomaly, a canonical example of global anomalies.
In contrast to the chiral anomaly, the parity anomaly induces the anomalous current \eqref{eq: anomalous current} with vanishing divergence, which implies it does not affect the second law of local thermodynamics in the absence of the boundary.

On the other hand, in the presence of the boundary, we showed the existence of the nontrivial edge currents combining the anomalous bulk current and the local second law.
Notably, the derived transport phenomena involve not only the $\U(1)$ current \eqref{eq: boundary particle current} but also the entropy current \eqref{eq: boundary entropy current} localized at the boundary.
Hence, our hydrodynamic equations clarify the nontrivial relation between the parity anomaly and the thermodynamic transports.
Besides, we also generalized our discussions on the parity anomaly to systems with the mixed anomaly among multiple $\U(1)$ and $\Z_2$ symmetries.

As a future study, it is worthwhile to identify the undetermined constant $c$ in the edge currents. 
The transport associated with $c$ is analogous to the energy current (or momentum density) of $(1+1)$-dimensional Weyl fermion~\cite{Kane:1996ads,Kane:1996bje,Cappelli:2001mp,Loganayagam:2012pz,Stone2012,Jensen:2012kj,Golkar:2015oxw,Chowdhury:2016cmh,Hongo:2019rbd}.
This becomes manifest when we move from the Landau frame to another frame by redefining the fluid velocity as%
\footnote{See e.g., Refs.~\cite{Tsumura:2012ss, Hayata:2015lga, Kovtun:2019hdm} for recent discussions about definitions of the fluid velocity in different frames in relativistic hydrodynamics.}
\begin{equation}
	u^\mu(x) \ra u^\mu(x) - \frac{\frac{C\mu(x)^2}{2}+cT(x)^2}{e(x)+p(x)}\epsilon^{\mu\nu\rho} \partial_\nu \Theta(\bx\in\Sigma)u_\rho(x).
\end{equation}
In this new frame, the part of the anomalous edge current \eqref{eq: boundary particle current} is eliminated. Instead, we find the following anomalous energy current at the boundary:
\begin{equation}
 \begin{split}
  T^{\mu\lambda}_\mathrm{edge}(x)
  = &- \left[\frac{C\mu(x)^2}{2}+cT(x)^2\right]
  \\
  &\times\left[\epsilon^{\mu\nu\rho} \partial_\nu \Theta(\bx\in\Sigma)u_\rho(x) u^\lambda(x)+(\mu\lra\lambda)\right].
 \end{split}
\end{equation}
The term involving $c$ is proportional to $T(x)^2$, which is an important feature of chiral transport in $(1+1)$ dimensions~\cite{Kane:1996ads,Kane:1996bje,Cappelli:2001mp,Loganayagam:2012pz,Stone2012,Jensen:2012kj,Golkar:2015oxw,Chowdhury:2016cmh,Hongo:2019rbd}.

As discussed in the last paragraph of Sec.~\ref{sec: boundary}, it is natural that the edge currents in our constitutive relations are half of the chiral transports of the $(1+1)$-dimensional Weyl fermion. 
Therefore, it can be reasonable to speculate $c = -\pi/24$, which is half of 
the coefficient of the energy current~\cite{Kane:1996ads,Kane:1996bje,Cappelli:2001mp,Loganayagam:2012pz,Stone2012,Jensen:2012kj,Golkar:2015oxw,Chowdhury:2016cmh,Hongo:2019rbd}.
One can perhaps figure out this coefficient by combining the partition function analysis~\cite{Banerjee:2012iz,Jensen:2012jh,Jensen:2012jy,Jensen:2013kka,Hongo:2016mqm,Manes:2018llx,Manes:2019fyw} and the global anomaly matching in the thermal spacetime~\cite{Jensen:2012kj,Golkar:2015oxw,Chowdhury:2016cmh,Glorioso:2017lcn,Hongo:2019rbd}.

Another interesting direction is to explore the possible anomalous edge current for systems with various anomalies in other dimensions.
From this viewpoint, the chiral anomaly may also give a nontrivial example
since the CME and CVE induce a seemingly dangerous term for the local second law at the boundary.
We expect that generalizing our formulation of hydrodynamics in the presence of the boundary can shed light on nontrivial dissipationless transports associated with the 't Hooft anomaly.

\acknowledgments
We thank Yuya Tanizaki for a variable discussion. TF thanks Yusuke Nishida for useful comments on the manuscript. TF was supported by JSPS KAKENHI Grants No.\ JP20J13415.
MH was supported by the US Department of Energy, Office of Science, Office of Nuclear Physics under Award No. DE-FG0201ER41195. 
This work was partially supported by the RIKEN iTHEMS Program (in particular iTHEMS Non-Equilibrium Working Group).

\bibliography{anomalymatching.bib}

\end{document}